\pdfoutput=1
\documentclass[aps,prl,amsmath, amssymb, english, twocolumn,reprint,floatfix, superscriptaddress]{revtex4-2}

\usepackage{multirow}
\usepackage{color}
\usepackage{float}
\usepackage[normalem]{ulem}
\usepackage{textgreek}
\usepackage{bbm}
\usepackage{comment}
\usepackage{soul}
\usepackage[unicode]{hyperref}
\hypersetup{
 colorlinks,
 citecolor=blue,
 filecolor=blue,
 linkcolor=blue,
 urlcolor=blue
}
\usepackage{microtype}
\usepackage{tikz}
\usepackage[applemac]{inputenx} 
\usepackage{amsmath}

\newcommand{\ket}[1]{|#1\rangle}

\begin{document}
\date{}
\title{Robust measurements of n-point correlation functions of driven-dissipative quantum systems on a digital quantum computer}
\author{Lorenzo Del Re} 
\affiliation{Department of Physics, Georgetown University, 37th and O Sts., NW, Washington,
DC 20057, USA}
\affiliation{Max Planck Institute for Solid State Research, D-70569 Stuttgart, Germany}
\author{Brian Rost} 
\affiliation{Department of Physics, Georgetown University, 37th and O Sts., NW, Washington,
DC 20057, USA}
\author{Michael Foss-Feig} 
\affiliation{Quantinuum, 303 S. Technology Ct, Broomfield, Colorado 80021, USA}
\author{A. F. Kemper} 
\affiliation{Department of Physics, North Carolina State University, Raleigh, North Carolina 27695, USA}
\author{J. K. Freericks} 
\affiliation{Department of Physics, Georgetown University, 37th and O Sts., NW, Washington,
DC 20057, USA}
\date{\today} 


\begin{abstract}
We propose and demonstrate a unified hierarchical method to measure $n$-point correlation functions that can be applied to driven, dissipative, or otherwise open or non-equilibrium quantum systems. In this method, the time evolution of the system is repeatedly interrupted by interacting an ancilla qubit with the system through a controlled operation, and measuring the ancilla immediately afterwards. We discuss the robustness of this method as compared to other ancilla-based interferometric techniques (such as the Hadamard test), and highlight its advantages for near-term quantum simulations of open quantum systems. We implement the method on a quantum computer in order to measure single-particle Green's functions of a driven-dissipative fermionic system. This work shows that dynamical correlation functions for driven-dissipative systems can be robustly measured with near-term quantum computers.
\end{abstract}
\maketitle

\emph{Introduction}.---
Open quantum systems, and in particular driven-dissipative systems, are among the most difficult problems to study in many-body physics, but also among the richest. The problem parameter space is vast; the bath as well as the system have their own inherent dynamics, and their interaction can be complex. Yet, in some sense there is a unification and emergent simplicity as the details often do not play a role when it comes to describing non-equilibrium steady (or periodic) states. These can be captured with a few parameters, have lost all knowledge of their history, and are stable to perturbations away from their fixed point. In other words, they are remarkably robust. 

Their robustness has also recently been exploited in simulations on quantum computers,
 either relying the hardware intrinsic decoherence  \cite{tseng2000,rost2020,Sommer2021},  by implementing Kraus maps and Lindblad operators  \cite{Barreiro2011,childs2016,cleve2016,delre2020,tornow2022non,hu2020,rost2021,hu2022general,PSchlimgen2021,Kamakari2022,Cattaneo2023}, or by implementing non-Hermitian dynamics \cite{Hubisz2021,zheng2021}.
In some cases, the existence of a fixed point has enabled quantum computers to perform the simulations far beyond the short coherence time of the qubits when the fidelity of one Trotter step is sufficiently high.
It thus appears that driven-dissipative systems are promising problems for applications of near-term quantum computers.

\begin{figure}
    \centering
    \includegraphics[width=0.95\columnwidth]{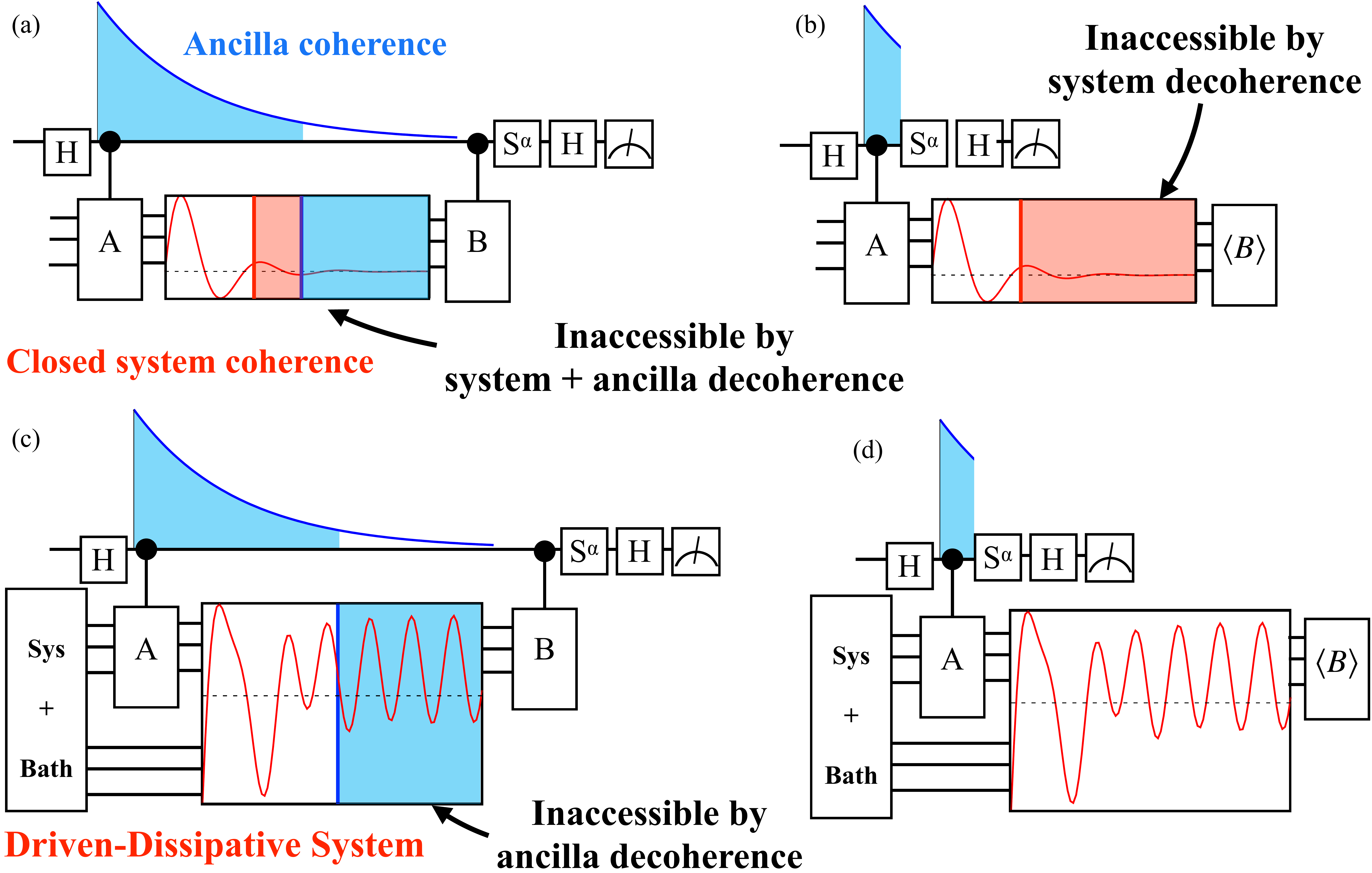}
    \caption{Overview of the proposed new method and its region of applicability. Ancilla decoherence (blue) limits the region where the Hadamard test method (a,c) yields a result, whereas the reset-based method (b,d) has no such limitation. For closed system evolution (a,b) the system decoherence naturally limits the maximal time of interest, but for driven-dissipative systems (c,d) the reset method is necessary to go beyond the ancilla coherence time.}
    \label{fig:cartoon}
\end{figure}

Given this situation, it is critical to develop the tools and methodology to be able to interrogate these long-lived non-equilibrium states.  For single-time operators this is not a problem,
one simply measures the desired operator at some point during the evolution. However, there is a wide class of observables --- correlation functions --- that are of equal or greater importance as they describe the excitations of the system, and make a direct connection to experimental observables.  The problem is that the typical protocols for the measurement of correlation functions 
\cite{Wecker2015,Keen_2020,steckmann2023mapping,greenediniz2023quantum,bishop2023quantum,dhawan2023quantum} are based on the Hadamard test \cite{Somma2002} [see Fig. \ref{fig:cartoon}(a)], where the correlation function is
measured with an ancilla qubit\cite{kreula2016few,kreula2016non,chiesa2019quantum,francis2020quantum,Jaderberg2020,sun2021quantum,steckmann2023mapping}.

This approach does not robustly generalize to measuring correlation functions in open quantum systems, and we illustrate the issues with it in Fig.~\ref{fig:cartoon} for a $2$-point correlation function. In short, the ancilla cannot capture the potentially long-time dynamics of the driven-dissipative open quantum systems because the ancilla has a short coherence time. For simulating closed quantum systems [Fig.~\ref{fig:cartoon}(a)] this is not a problem because the system has an equally short coherence time, and the region of inaccessibility by the Hadamard test approach has no information. On the other hand, for simulating open quantum systems
[Fig.~\ref{fig:cartoon}(c)], this is a problem because the now-stable dynamics of the driven-dissipative system are inaccessible due to the ancilla decoherence.

In this Letter, we propose a strategy that (i) is a full framework for computing $n$-point correlation functions in open quantum systems, and (ii) is suitable for the near term where we cannot rely on long ancilla coherence lifetimes. Its crux is a measurement of the ancilla right after it is entangled with the system, and using the result of the measurement in post-processing to construct the desired $n$-point correlation function. We illustrate the idea in Fig.~\ref{fig:cartoon} (b) and (d).  For simulating closed quantum systems this approach yields the same information as the standard Hadamard test; however, for open quantum systems the region of inaccessibility by ancilla decoherence is now fully accessible.

Our method is a simple strategy capable of measuring arbitrary unequal-time correlation functions between multi-qubit Pauli operators, and which works for both dissipative and unitary time evolution. As such, it subsumes and unifies the approaches of Ref. \cite{Knap2013} (unequal-time commutators) and Ref. \cite{Uhrich2017} (unequal-time anticommutators).
It is hierarchical in the sense that extracting the information of an $n$-th order correlation function requires previous knowledge of lower-order correlation functions; but, it restores the robust nature of driven-dissipative systems, because it does not require system-ancilla entanglement to be maintained during the time evolution of the system. We verify the validity of our method by performing measurements of the single-particle Green's function of a driven-dissipative fermionic model using a Quantinuum quantum computer. Our results show excellent quantitative agreement between  data and  the theoretical predictions.

 \par 
 \emph{Target quantities.}--- Our goal is the calculation of correlation functions of a generic system (S) that can also dissipate energy through an interaction with a bath (E); so we employ the density matrix formalism, which is required to study open quantum systems \cite{Breuer2002}.
 \par The correlation functions
 are constructed as follows.
 Let $\left\{O_i\right\}$ be a set of operators in the Schr\"{o}dinger representation acting on the system Hilbert space with $i = 1,2,\cdots n$, and let $\{t_i\}$ being a set of ordered time values such that $t_0<t_1<t_2< \hdots t_{n-1}<t_n$, where $t_0$  is the initial time, then we define the $n$-th rank correlation function via
 \begin{align}\label{eq:corr}
 &\left<O_{n}(t_{n})O_{{n-1}}(t_{{n-1}})... O_{1}(t_{1})\right>\nonumber\\
 &~~~~~=\mbox{Tr}_\text{S}\left\{O_n\mathcal{V}_{t_n,t_{n-1}}... O_2\mathcal{V}_{t_2,t_1}O_1\mathcal{V}_{t_1,t_0}\rho(t_0)\right\}.
  \end{align}
Here, $O_i(t_i)$ is the operator $O_i$ in the Heisenberg representation,
$\rho(t_0)$ is the system density matrix evaluated at the initial time, $\mathcal{V}_{t_{i+1},t_i}$ is the time evolution super-operator that evolves the system from time $t_i$ to $t_{i+1}$ (i.e. $\rho(t_{i+1}) = \mathcal{V}_{t_{i+1,t_i}}\rho(t_i)$ acting from left to right), and $\mbox{Tr}_\text{S}$ indicates a trace over the system subspace (meaning that the degrees of freedom of the bath have already been integrated out).
 For simplicity, and without loss of generality, we assume that the operators $O_i$ are unitary  and Hermitian operators; addressing this case is sufficient to demonstrate the validity of our method, because a non-unitary operator can always be  expanded as a linear combination of unitaries, chosen to also be Hermitian  (e.g. Pauli strings).
 As will be shown in the next section,  the correlation function in Eq.~(\ref{eq:corr}) can be extracted from the Hadamard test.
 \par
\begin{figure*}[btph]
    \centering
    \includegraphics[width = 1.8\columnwidth]{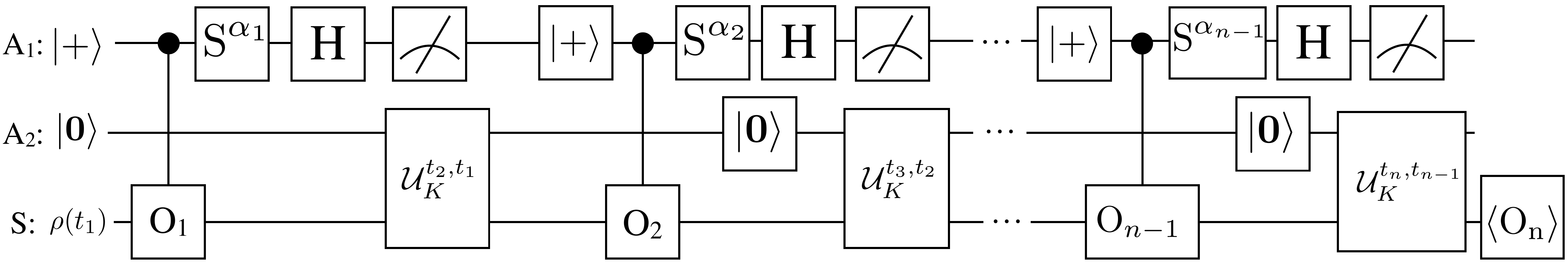}
    \caption{Circuit to measure a generic $n$-time correlation function of the kind defined in Eq.(\ref{eq:corr_comm}) using the robust strategy.}
    \label{fig:alt_sch}
\end{figure*}

 The alternative strategy that we propose  will naturally yield correlation functions of nested commutators and anti-commutators of the form
 \begin{align}\label{eq:corr_comm}
   \!\! \Big\langle[O_1(t_1),\!\big[O_2(t_2),\dots\,\!\big[O_{n-1}(t_{n-1}),O_n(t_n)\big]_{\pm}\!\,\cdots\!\,\big]_{\pm}\big]_{\pm}\Big\rangle,
 \end{align}
 where $[.\,,.]_{\pm}$ can be either  commutators (-) or  anti-commutators (+), all chosen independently.  The correlation function in Eq.~(\ref{eq:corr}) can be obtained from the one in Eq.~(\ref{eq:corr_comm}) and vice versa by performing multiple measurements and then combining the different outcomes together. \footnote{The correlation function in Eq.~(\ref{eq:corr_comm}) is not the most general $n$-point correlation function, because it always maintains Keldysh time ordering on a two branch Keldysh contour, as opposed to the most general form, which requires higher-order Keldysh contours~\cite{Tsuji2017}; for concrete examples, see the supplemental information \cite{supp}}.  We note that in the case of two-point functions, Eq.~(\ref{eq:corr}) corresponds to lesser or greater  Green's functions while Eq.~(\ref{eq:corr_comm}) to advanced, retarded, and Keldysh Green's functions \cite{stefanucci2013}, so both methods produce all the physical Green's functions needed to describe a time evolving quantum system. However, there are some limitations --- for example
  one cannot directly calculate out-of-time-ordered correlation functions with the circuit in Fig.~(\ref{fig:alt_sch}) and we leave possible generalizations of this method to future work. 
 \\
 \par
 \begin{figure}[htpb]
     \centering
     \includegraphics[width = 0.95\columnwidth]{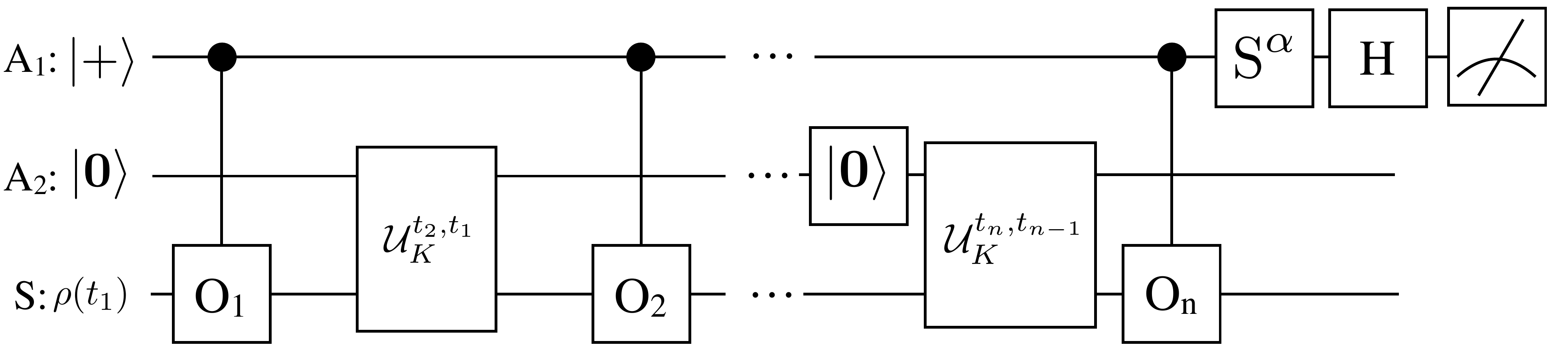}
     \caption{The standard interferometric scheme for measuring  the $n$-time correlation function~\protect{\cite{Somma2002}}, as given in Eq.~(\ref{eq:corr}), for a dissipative circuit.  Accurate results require that the ancilla register $A_1$ maintain coherence over the entire duration of the circuit. }
     \label{fig:Rams}
 \end{figure}
 \emph{Hadamard test for driven-dissipative systems.}---
 In Fig.~\ref{fig:Rams}, we show how the interferometry scheme proposed in  Ref.~\onlinecite{Somma2002} generalizes to compute the $n$-time correlator defined in Eq.~(\ref{eq:corr}) for an open quantum system. In order to simulate dissipative dynamics, we need a generic $k$-qubit ancilla register  (called A$_2$) that we take to be initialized into the state $\left|\boldsymbol{0}\right\rangle = \ket{0}^{\otimes k}$. 
 A suitable unitary operation $\mathcal{U}_K^{t,t^\prime}$ that entangles A$_2$ with the system register S followed by tracing out (ignoring) the state of the ancilla register can encode
 the non-unitary time evolution map $\mathcal{V}_{t,t^\prime}$, which can be rewritten using the Kraus sum representation:
 \begin{equation}\label{eq:Kraus}
     \mathcal{V}_{t,t^\prime} \,\rho(t^\prime) = \sum_{i=0}^{2^k-1}K^{\,}_i(t,t^\prime)\rho(t^\prime)K^\dag_{i}(t,t^\prime),
 \end{equation}
 where $K_i$ are the so called Kraus operators satisfying the sum rule $\sum_i K^\dag_i(t,t^\prime)K^{\,}_i(t,t^\prime) = \mathbbm{I}$. They are related to the unitary evolution of the system and ancilla bank  by $K_i(t,t^\prime) = \left\langle i\right|\mathcal{U}_K^{t,t^\prime}\left|\boldsymbol{0}\right\rangle$, with $\{\left|i\right\rangle\} $ being a complete basis for A$_2$. In the interferometry scheme, we need an extra single-qubit ancilla register A$_1$  in which all the information about the correlation function (which is a complex number) will be stored. For example, in the case of $n = 2$, the final quantum state of the A$_1$ qubit reads:
\begin{align}\label{eq:final_rams}
     \rho_{A_1} &=\left\{
     \begin{array}{c}
     \frac{1}{2} \hat{\mathbbm{I}}_{A_1} + \frac{1}{2}\text{Re}[C]\,\hat{\sigma}^z_{A_1} -\frac{1}{2}\text{Im}[C] \hat{\sigma}^y_{A_1}\,, \text{if } \alpha = 0, 
     \\ \\
     \frac{1}{2} \hat{\mathbbm{I}}_{A_1} - \frac{1}{2}\text{Im}[C]\,\hat{\sigma}^z_{A_1} -\frac{1}{2}\text{Re}[C] \hat{\sigma}^y_{A_1}\,, \text{if } \alpha = 1, 
     \end{array}
     \right.
 \end{align}
 where $C = \left<O_2(t_2)O_1(t_1)\right>$, and the binary variable $\alpha=\{0,1\}$ indicates whether
 the $S$ gate was applied or not.
Measuring the ancilla in the $Z$ and $Y$ bases determines the real and imaginary parts of the correlation function.
 \par This method is convenient because the complex information encoded in the correlation functions of a many-body system are found from single qubit measurements. However, this scheme requires maintaining the coherence of the A$_1$ ancilla (and thereby its entanglement with the system) for the full duration $t_n-t_1$. In the next section, we introduce an alternative robust scheme that does not require maintaining coherence of the $A_1$ ancilla, but at the cost of requiring a more complex measurement scheme.  
 \\
 \par
 
 \emph{Robust strategy.}---
In Fig.~\ref{fig:alt_sch}, we show the alternative circuit to measure the correlation function defined in Eq.~(\ref{eq:corr}). This circuit is schematic, because it encodes all possible circuits that are employed to measure the set of correlation functions in Eq.~(\ref{eq:corr_comm}). Here,  each realization has  chosen unitary operations acting on A$_1$ (selected from $(S)^{\alpha_i}H  $, where $S$ and $H$ are the  phase gate and the Hadamard gate, respectively) 
for each time $t_i$ measured in the correlation function. 
The circuit shown in Fig.~\ref{fig:alt_sch} naturally measures the set of correlation functions defined in Eq.~(\ref{eq:corr_comm}) with the commutator or anti-commutator chosen from the $n-1$ dimensional binary vector $\boldsymbol{\alpha}= \{\alpha_1,\alpha_2,...,\alpha_{n-1}\}$.
 It is important to note that after the $S^{\alpha_i}H$  operation is performed, the ancilla qubit A$_1$ is measured immediately afterwards and the measurement outcome ($m_i$) is stored; such a measurement destroys the entanglement between A$_1$ and the state encoded in the system and the A$_2$ ancilla bank.
 The state is then evolved to the next $t_i$  using the Kraus map decomposition defined in Eq.~(\ref{eq:Kraus}). The $A_1$ ancilla is then reset to its $\ket{+}$ state and the process is repeated for each operator in the correlation function.  In the last step, after the final time evolution from $t_{n-1}$ to $t_n$, the $S$ register qubits will be in a final state $\rho_n$ and the  operator O$_n$ is measured directly on the $S$ register qubits, yielding results that depend on $\boldsymbol{\alpha}$. The correlation function is determined by classical post-processing of the accumulated results and the choice of $\boldsymbol{\alpha}$.
 \par In general, the state of the system qubits at time $t_{j+1}$ is obtained from the state at $t_{j}$ through the following map \cite{supp}: 

\begin{align}\label{eq:two_point}
    &\rho_{j+1} \propto \mathcal{V}_{t_{j+1},t_{j}}\left(\rho_j+O_j\rho_jO_j+[(-1)^{m_j}i^{\alpha_j}O_j\rho_j+\text{h.c.}]\right),
\end{align}
where the proportionality constant is given by tracing the RHS of the equation. Here, $m_j = \{0,1\}$ is the result of the A$_1$ qubit measurement,  and $\rho_{j = 1} = \rho(t_1)$ is given by the initial state of the system at time $t_1$ (see Fig.~\ref{fig:alt_sch}). 
\par In order to show how this method works in practice, we discuss the two simplest cases: i.e. the two-point and the three-point correlation functions.
For $n = 2$, 
 the result of measuring $O_2$  directly on the system register will yield 
\begin{align}\label{eq:two_point_av}
   &\text{Tr}\, O_2\,\rho_2 =\mathcal{N}\left\{ \left<O_2(t_2)\right> + \left<O_2(t_2)\right>_{O_1} \right. \nonumber \\
    &+\underbrace{\left.{(-1)^{m_1}}\left[i^{\alpha_1}\left<O_{2}(t_2)O_1(t_1)\right>  -i^{-\alpha_1}\left<O_{1}(t_1)O_2(t_2)\right>\right]\right\}}_{\propto  \left<[O_1(t_1),O_2(t_2)]_\mp\right>},  \nonumber \\
\end{align}
where $\mathcal{N} = \{2 +\big[ (-1)^{m_1}i^{\alpha_1}\left<O_1(t_1)\right> + \text{h.c.}\big]\}^{-1}$,  $\left<O_1(t_1)O_2(t_2)\right> = \text{Tr}\left(O_2\,\mathcal{V}_{t_2,t_1}[ \rho(t_1)O_1]\right)$, and $\left<O_2(t_2)\right>_{O_1} :=  \text{Tr}\left(O_2\,\mathcal{V}_{t_2,t_1}[O_1\rho(t_1)O_1]\right)$. 
Hence, when $\alpha_1 =0, 1$ the term in square brackets in Eq.~(\ref{eq:two_point_av}) is proportional  to $\left<[O_1(t_1),O_2(t_2)]_\mp\right>$. This is precisely Eq.~(\ref{eq:corr_comm}) when $n = 2$. 
 In order to isolate our target quantity, i.e. the two-time correlation function $\left<[O_1(t_1),O_2(t_2)]_\mp\right>$, we need to subtract the first two terms in Eq.~(\ref{eq:two_point_av}) multiplied by $\mathcal{N}$  which are determined by equal-time averages that can be obtained by performing a set of simple extra measurements \cite{supp}. It is worthwhile to note, that the measurement outcome of the ancilla  $m_1 =0,1$ must be stored for extracting our target quantity.  

\par For $n = 3$, measuring $O_3^{\,}$  results in the following quantity:
\begin{eqnarray}\label{eq:three_point}
     \text{Tr}\,O_3\,\rho_3 =  (-1)^{m_1+m_2}\,C^{\boldsymbol{\alpha}}_{t_1,t_2,t_3} + R_{\boldsymbol{\alpha}},
\end{eqnarray}
where   $C^{\boldsymbol{\alpha}}_{t_1,t_2,t_3}$ is a three-time correlation function that depends on the values of $\boldsymbol{\alpha}$. There are four possible values    $\left<[O_1^{\,}(t_1),[O_2{\,}(t_2),O_3^{\,}(t_3)]_\pm]_{\pm}\right>$. In addition, there are contributions denoted by $R_{\boldsymbol{\alpha}}$, which is a  remainder function. It is determined by performing additional measurements  comparable to what is needed for lower-rank correlation functions (see supplemental information for details: \cite{supp} ).
\par 
 We note that in the case of  single-qubit \cite{Knap2013,Uhrich2017,Schuckert2020} and two-qubit \cite{mitrarai2019} correlators, there are alternative ways of measuring correlation functions that do not require the extra ancilla register  A$_1$, {at the cost of performing more measurements on the system \cite{supp}}.
 \par 
%
\begin{figure}[htpb]
     \centering
     \includegraphics[width=\columnwidth]{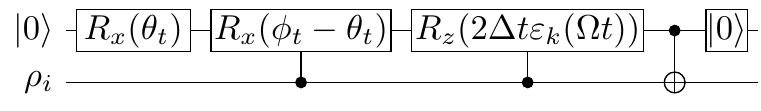}
     \caption{Circuit implementing the Trotterised time evolution $\mathcal{U}_K(t,t+\Delta t)$ of the model defined in Eq.~(\ref{eq:me_lin}).}
     \label{fig:trott_dyn}
 \end{figure}
 \emph{Hardware implementation}--- In order to verify the validity of the protocol, we applied it to measure the Green's function of spinless
 free fermions in a lattice driven by a constant electric field that also dissipate energy through a coupling with a thermal bath. 
 The  Hamiltonian of this chosen system plus bath can be brought into a block-diagonal form after performing a Fourier transform  to momentum 
 space as described in Ref.~\cite{delre2020}. Hence, the system's 
 reduced density matrix 
 factorizes
as a tensor product in momentum space, i.e., $\bigotimes_k\rho_k$
, and  we can define a (diagonal in $k$) master
equation for each 2$\times 2$ $k$-dependent density matrix $\rho_k$,
\begin{equation}\label{eq:me_lin}
    \partial_t \rho_k = -i[\mathcal{H}_k(t),\rho_k]+\sum_{\ell=\{1,2\}}L_{\ell}^{\,} \,\rho_k \,L_\ell^{\dag} - \{\rho_k,L^\dag_\ell L^{\,}_\ell\},
\end{equation}
where the Lindblad operators are $L_1 = \sqrt{\Gamma\,n_F(\epsilon_k(t))}d_k$ and $L_2 = \sqrt{\Gamma\,n_F(-\epsilon_k(t))}d^\dag_k$, with $d_k^{\,}$ being the destruction operator of a lattice fermion with quasi-momentum $k$, $\epsilon_k(t) = - 2J \cos\left(k+\Omega\,t\right)$, with $J$ being the hopping amplitude, $\Omega$ the amplitude of the applied DC field, and $k$ the crystalline momentum.   $\Gamma$ sets the strength of the system-environment coupling and $n_F(x) = \left[1 + \exp(\beta\, x)\right]^{-1}$ is the Fermi-Dirac distribution with $\beta$ being the inverse of the bath temperature.
In Fig.~\ref{fig:trott_dyn}, we show the circuit implementing $\mathcal{U}_K$ for the Kraus map related to Eq.~(\ref{eq:me_lin}).
The Lindblad operator $L_{1(2)}$ encodes the physical process of a Bloch electron (hole) with momentum $k$ to hop from the lattice to the bath with a probability given by $\Gamma n_F[-\epsilon_k(t)]$ $(\Gamma n_F[\epsilon_k(t)])$. Such a decay process introduces a time dependence of the momentum distribution function of fermions and a damping of Bloch oscillations that eventually leads to a non-zero average of the DC-current \cite{supp,han2013,delre2020,rost2021}. 
\begin{figure}[bhtp]
    \centering
    \includegraphics[width = 0.95\columnwidth]{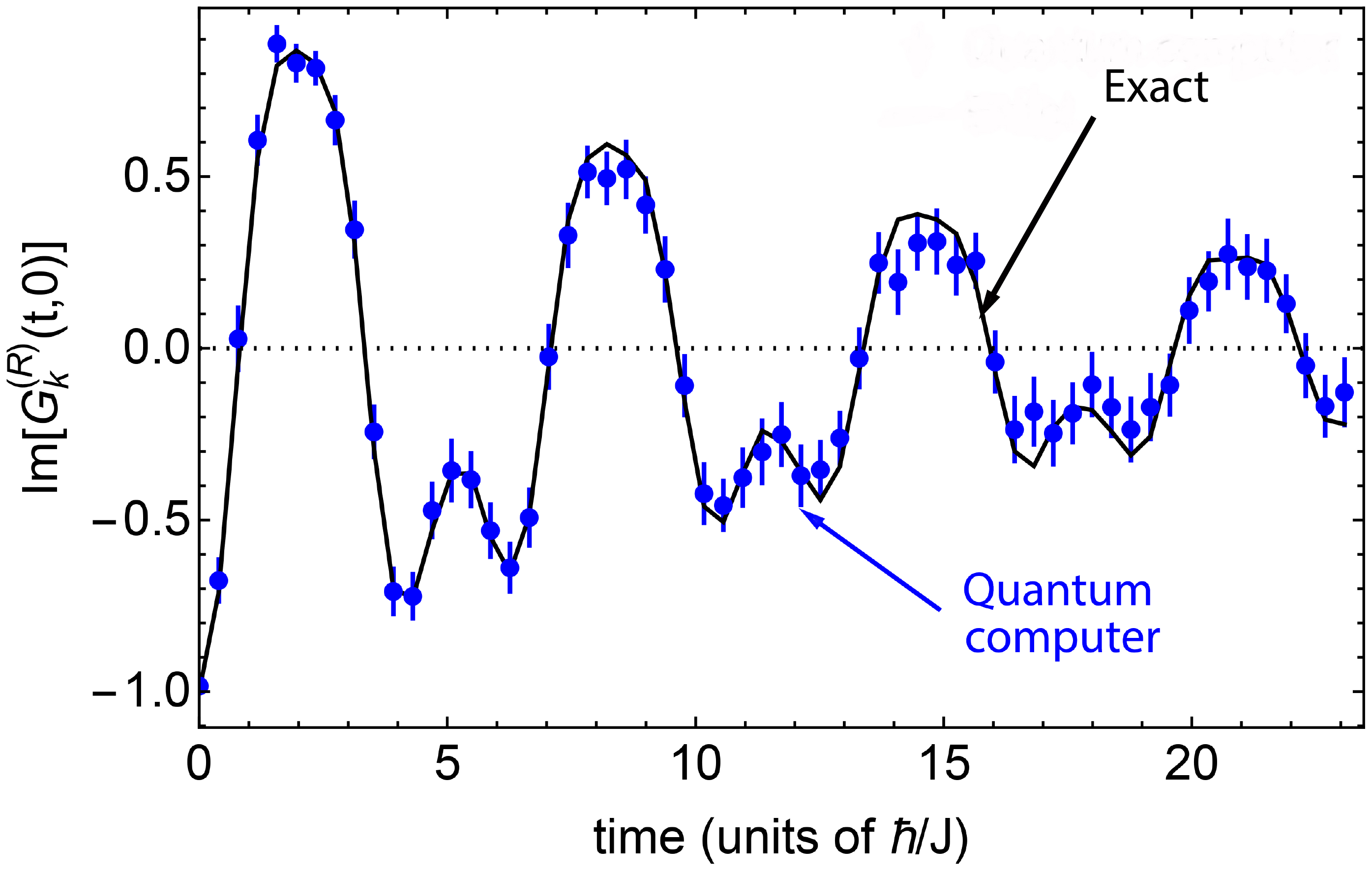}
     \includegraphics[width = 0.95\columnwidth]{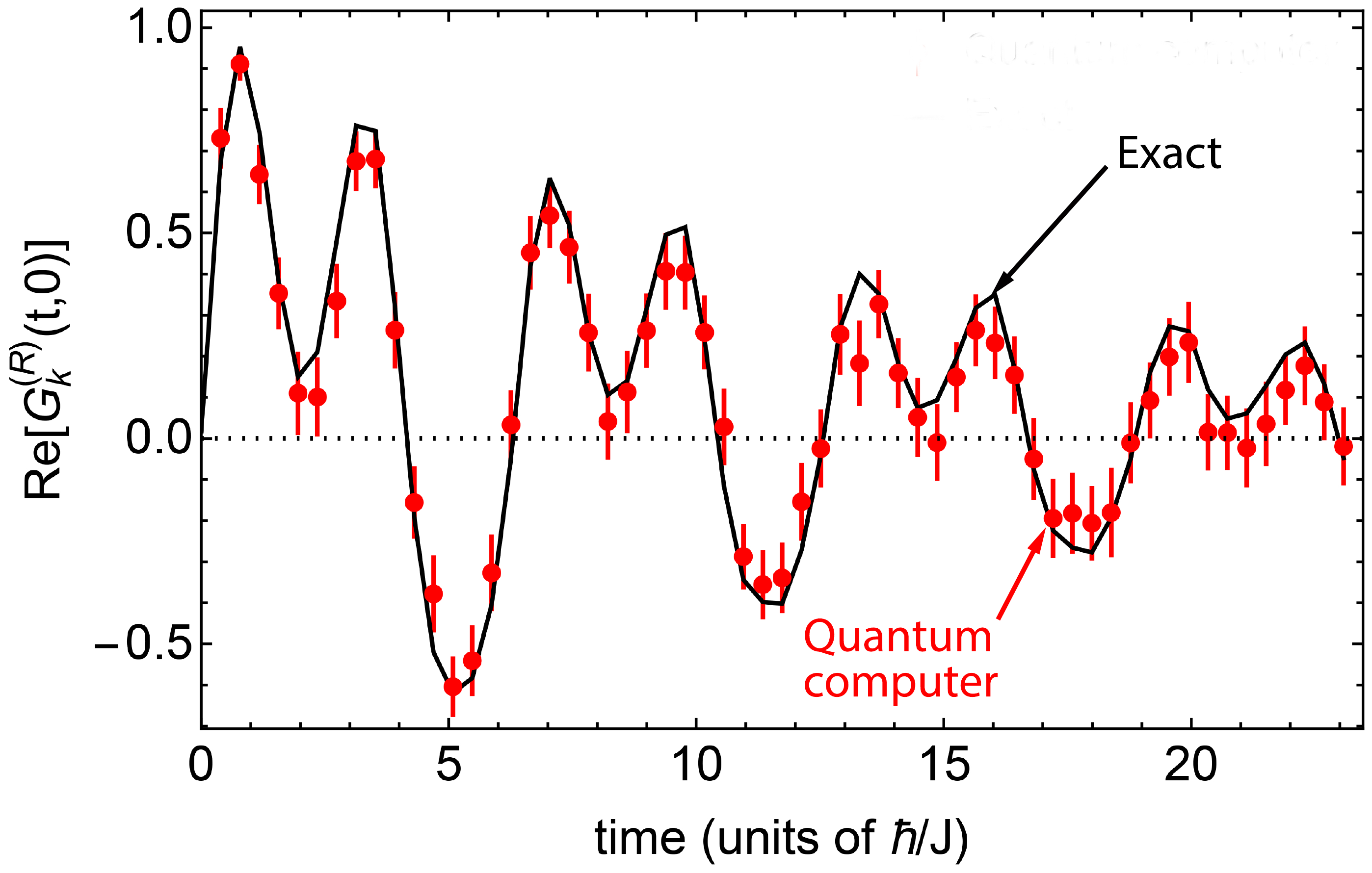}
    \caption{ Imaginary (upper panel) and real (lower panel) parts of the retarded fermion Green's function as a function of time (parameters are wavevector $k=-0.5/a_0$ where $a_0$ is the lattice constant, dimensionless electric field is $\Omega=1$, the dissipation rate to the fermionic bath is $\Gamma=1/16$ and the bath temperature is 0.01 in units of the hopping). Circles represent data from a Quantinuum model-H1-2 quantum computer, with error bars representing 2$\sigma$ confidence intervals.  The primary source of error in the implemented circuits is due to noise on the two-qubit gates [$(2-3)\times 10^{-3}$ average infidelity]. However, the resulting noise model leads to results that are barely distinguishable by eye from exact circuit simulations (black lines) and are omitted from the figure.}
    \label{fig:QC_res}
\end{figure}
 \par 
 In Fig.~\ref{fig:QC_res}, we show the retarded fermion Green's function $G_k^{(R)}(t,t^\prime) = -i\theta(t-t^\prime)\left<[d^{\,}_k(t),d^{\dag}_k(t^\prime)]_+\right>$ measured on Quantinuum's model H1 quantum computer.
 The retarded Green's function of the model can be computed exactly and its derivation and analytical form are given in the supplemental materials \cite{supp}.
 \par
 There is excellent quantitative agreement between the data produced by the quantum computer and the expected curves in presence of noise. It is worthwhile to note that in the presence of a driving field, the Green's function does not oscillate as a simple sinusoidal function and it presents extra features, such as the additional maxima and minima occurring between times 10 and time 19  [see Fig.~\ref{fig:QC_res}], that are faithfully reproduced by the quantum computer data. 
 \par 

 \emph{Conclusion \& Outlook.}--- We have put forward a robust technique for the measurement of multi-point correlation functions of driven-dissipative quantum systems that can be applied in the realm of quantum simulations of complex models such as the Hubbard model. 
 Unlike the Hadamard test, which requires us to keep the ancilla and system qubits coherently entangled, our new approach does not. This is advantageous for driven-dissipative systems, where the system is not coherent (see Fig.~\ref{fig:cartoon}), although it comes at the cost of performing extra measurements, as well as requiring additional circuits of lower depth than the one needed to extract the target quantity. Our method naturally computes correlators of the form given in Eq.~(\ref{eq:corr_comm}), which
 represent a myriad of response functions and experimental measurements.
 We applied our method to measuring the Green's function of free fermions driven out of equilibrium and interacting with a bath. The data obtained from the quantum computer are in an excellent agreement with the curves predicted by the theory. While this data constitutes an important proof of principle enabling the measurement of correlation functions on near-term quantum computers, further work needs to be done to use this approach to solve new problems in science. 
 \par Interestingly, given its generality, the Hadamard test has applications other than the measurement of correlation functions; for example, it has been proposed for determining important overlaps in the realm of variational quantum dynamics simulations \cite{Yuan2019,Yao2021} and also for the simulation of open quantum systems using quantum imaginary-time evolution \cite{Kamakari2022}. We therefore expect our robust alternative strategy to the Hadamard test to be suitable for these other applications as well.

 \emph{Acknowledgments.}--- We acknowledge financial support
from the U.S. Department
of Energy, Office of Science, Basic Energy Sciences, Division of Materials Sciences and Engineering. Initial submission and project execution were performed under Grant No.
DE-SC0019469; the resubmission under
Grant no. DE-SC0023231.
BR was also funded by the National Science Foundation under  Award No. DMR-1747426
(QISE-NET).
JKF was also funded by the McDevitt bequest at Georgetown University.
 \par 

\bibliography{ref_bibilio} 
\clearpage

  \end{document}


\date{}
\title{Supplemental material for: Robust measurements of n-point correlation functions of driven-dissipative quantum systems on a digital quantum computer}
\author{Lorenzo Del Re} 
\affiliation{Department of Physics, Georgetown University, 37th and O Sts., NW, Washington,
DC 20057, USA}
\affiliation{Max Planck Institute for Solid State Research, D-70569 Stuttgart, Germany}
\author{Brian Rost} 
\affiliation{Department of Physics, Georgetown University, 37th and O Sts., NW, Washington,
DC 20057, USA}
\author{Michael Foss-Feig} 
\affiliation{Quantinuum, 303 S. Technology Ct, Broomfield, Colorado 80021, USA}
\author{A. F. Kemper} 
\affiliation{Department of Physics, North Carolina State University, Raleigh, North Carolina 27695, USA}
\author{J. K. Freericks} 
\affiliation{Department of Physics, Georgetown University, 37th and O Sts., NW, Washington,
DC 20057, USA}
\date{\today} 

\maketitle

\renewcommand\theequation{S\arabic{equation}} 
\renewcommand\thefigure{S\arabic{figure}}  
\renewcommand\thetable{S\arabic{table}}  
\setcounter{figure}{0}

\section{Correlation functions on the Keldysh contour}
We discuss more in detail what kind of correlation functions can be evaluated by the circuit in Fig.~(3) in the main text and what are those for which a generalization of our scheme is needed.
In  general, since the operators $O_i(t_i)$ do not commute with each other,  one can construct $n!$ correlation function out of all possible permutations of the operators in Eq.~(1) in the main text.
However calculating all possible nested commutators/anti-commutators of Eq.(2) gives  the possibility of isolating only $2^{n-1}$ of these permutations. 
\par
\begin{figure}[hbtp!]
    \centering (a)
    \includegraphics[width = 0.9\columnwidth]{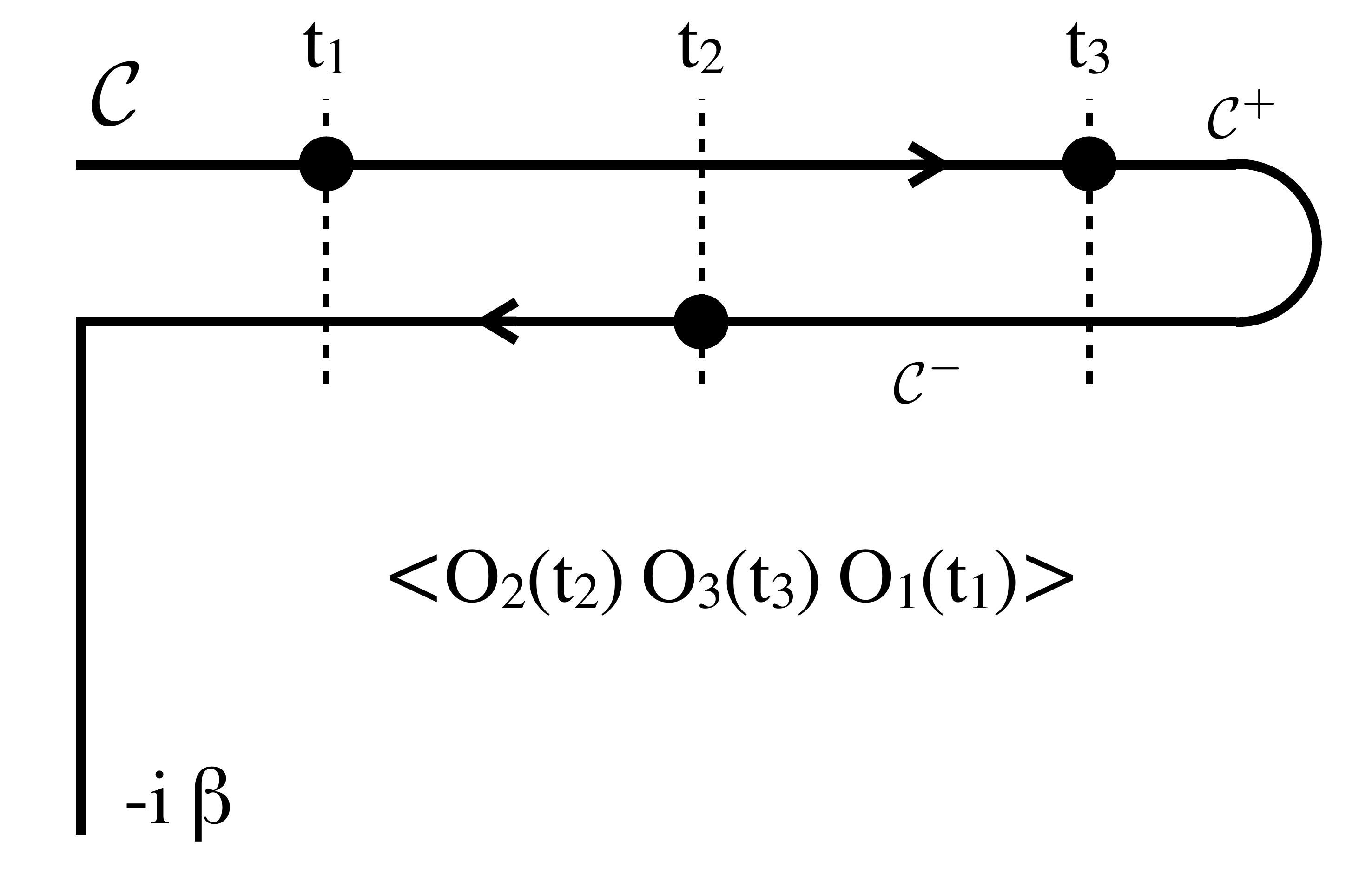}
    (b)
     \includegraphics[width = 0.85\columnwidth]{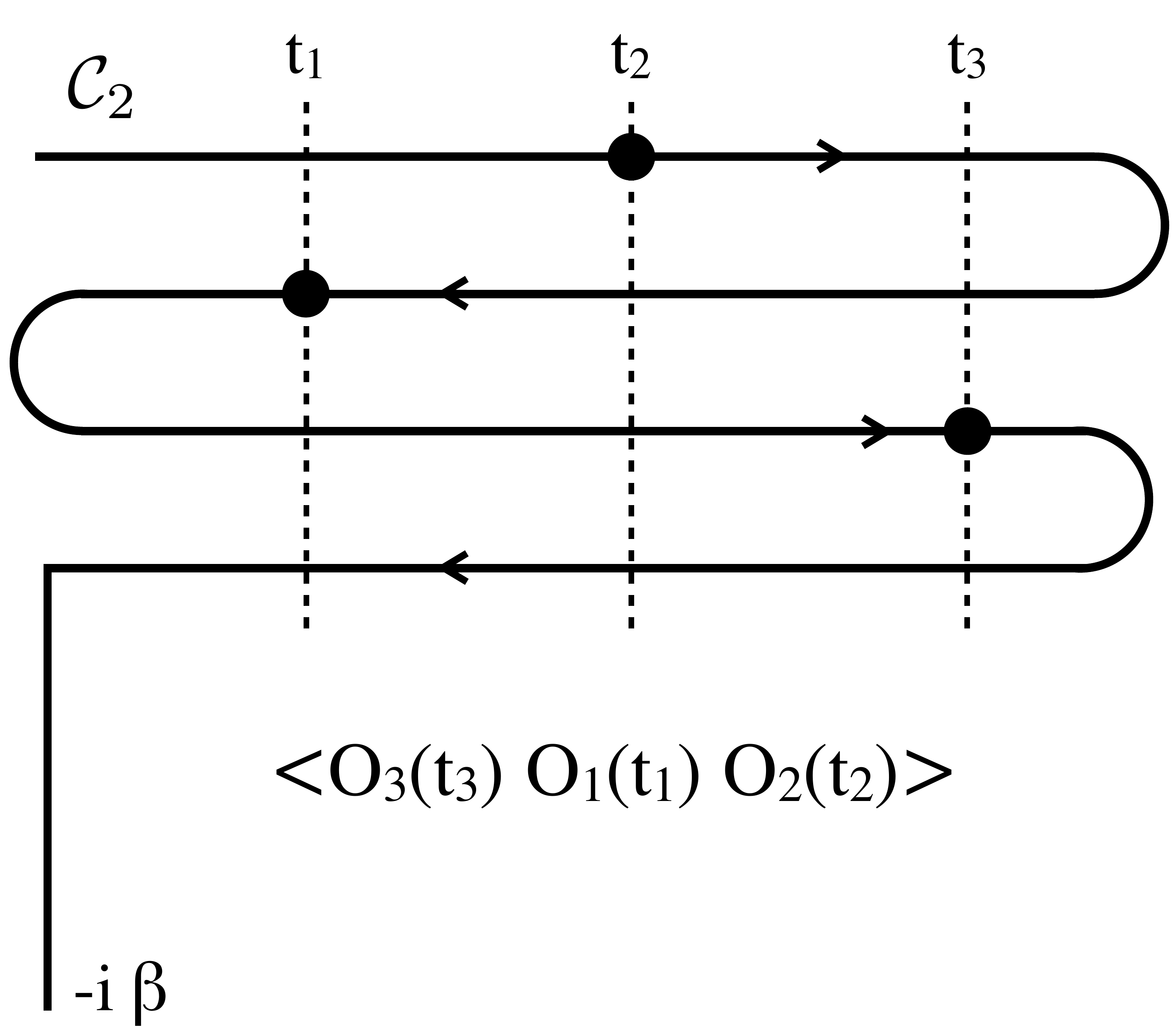}
    \caption{(a) Simple  Keldysh contour $\mathcal{C}$ made up of the two forward ($\mathcal{C}^+$) and backward ($\mathcal{C}^-$) branches. Black dots specify on which branch the three different operators appearing in the permutation $\left<O_2O_3O_1\right>$ are evaluated.  (b) Multiple-branch Keldysh contour $\mathcal{C}_2$
    needed for the evaluation of the permutation $\left<O_3O_1O_2\right>$. 
    In this case, one cannot evaluate $O_3$ 
    on one of the first two branches and keep the product $O_3 O_1 O_2$ ordered on the contour at the same time. Time-ordering along the Keldysh contour, regardless of the number of branches (or whether time increases or decreases), simply proceeds by ordering along the arc length of the contour from start to finish.}
    \label{fig:keldyshc}
\end{figure}

 Here we argue that the permutations we have access to correspond to components of time-ordered correlation functions on a simple two-branch Keldysh contour [see Fig.(\ref{fig:keldyshc}a) ], while the remaining ones are components of time-ordered correlation functions on a more complicated contour with multiple branches [see Fig.(\ref{fig:keldyshc}b)].
\par
Instead of giving a rigorous proof of our statement,  we will consider the specific example for $n = 3$. In this case, the  permutations we have access to through Eq.~(2) are:
$\left<O_1(t_1)O_2(t_2)O_3(t_3)\right>$, $\left<O_1(t_1)O_3(t_3)O_2(t_2)\right>$, $\left<O_2(t_2)O_3(t_3)O_1(t_1)\right>$ and $\left<O_3(t_3)O_2(t_2)O_1(t_1)\right>$, where $t_3> t_2>t_1$. Hence, the following two permutations $\left<O_3(t_3)O_1(t_1)O_2(t_2)\right>$ ,  $\left<O_2(t_2)O_1(t_1)O_3(t_3)\right>$  are missing.
\par In Fig.(\ref{fig:keldyshc}a) we show a two-branch Keldysh contour \cite{stefanucci2013} $\mathcal{C} = \mathcal{C}_+ + \mathcal{C}_-$, where $\mathcal{C}_\pm$ are the forward and backward branches. In the same figure we show that the particular permutation $\left<O_2O_3O_1\right>$ can be seen as a time-ordered function on the contour $\mathcal{C}$ where $t_1 = t_{1+}$ and $t_3 = t_{3+}$ lie on the forward branch while $t_2 = t_{2-}$ lie on the backward branch. 
\par Conversely, the permutation $\left<O_3O_1O_2\right>$ cannot be interpreted as a time-ordered correlation function on $\mathcal{C}$. Instead, it could be seen as a time-ordered correlation function on the more complex contour $\mathcal{C}_2$  shown in Fig.(\ref{fig:keldyshc}b) that contains two additional branches. The same contour has to be considered for the calculation of out-of-time correlators \cite{Tsuji2017}, and therefore a generalization of our scheme is needed in order to have access to those quantities. 
{
\section{Time step update}
\begin{figure}
    \centering
    \includegraphics[width = \columnwidth]{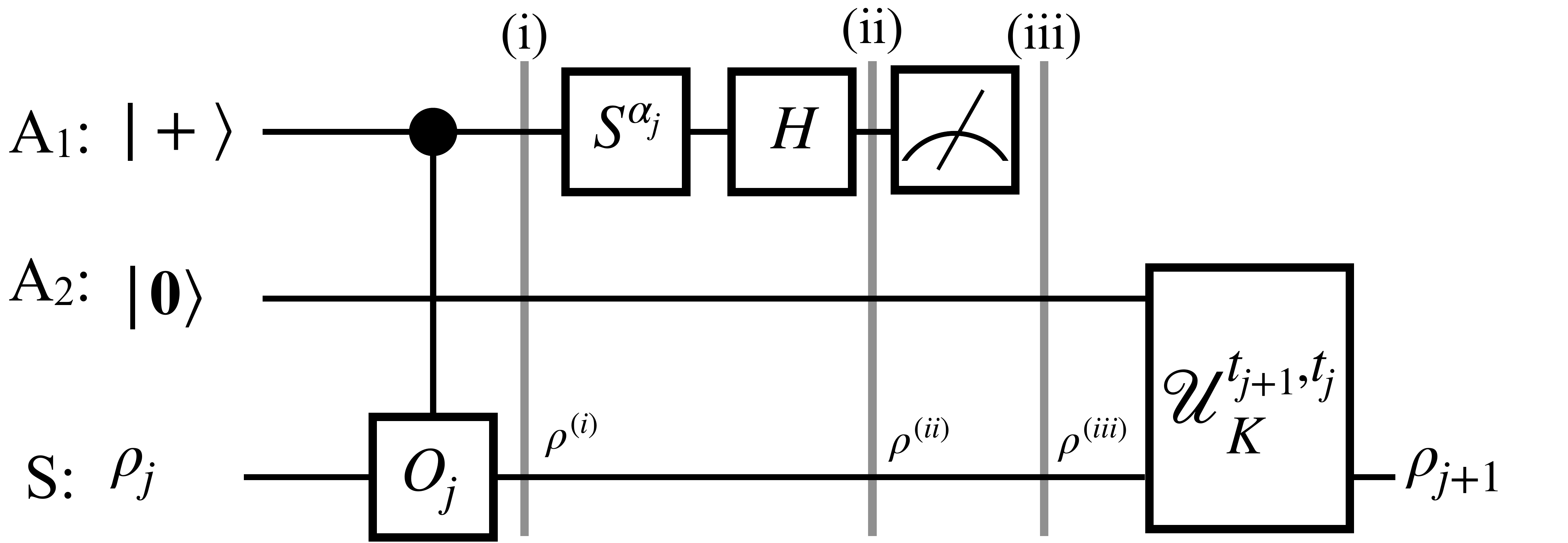}
    \caption{ Detail of the circuit presented in Figure 3 in the main text that is responsible for a single time step  evolution from time $t_j$ to time $t_{j+1}$.}
    \label{fig:single_step_sketch}
\end{figure}
In this section, we derive Eq.~(5) of the main text, which describes how the density matrix is updated from the $j$-th step  to  $(j+1)$-th one. In Figure \ref{fig:single_step_sketch}, we show the section of the circuit presented in Fig.~3 of the main text which accounts for the single step time evolution. 
We note that we have two ancilla registers $A_1$ and $A_2$. $A_2$ is a single ancilla register given by $k$-qubits needed to implement the dissipative time evolution (i.e. the Kraus map) of the system density matrix. This register is initialized in the computational basis as 
$\ket{0}^{\otimes k}$.
$A_1$ is a single-qubit ancilla register needed for the correlation function measurement protocol.
We marked three intermediate points in the circuits $(i)$, $(ii)$ and $(iii)$ which correspond respectively to the density matrices of system plus ancilla register $A_1$ $\rho^{(\text{i})}$, $\rho^{(\text{ii})}$ and  $\rho^{(\text{iii})}$. 

After the first control operation on the system we have: 
\begin{align}
    \rho^{(\text{i})} &= \rho_j \otimes \ket{0}\bra{0} +  O_j \rho_j \otimes \ket{1}\bra{0} \nonumber \\
    &+ \rho_j O_j  \otimes \ket{0}\bra{1}  + O_j \rho_j O_j  \otimes \ket{1}\bra{1}.
\end{align}
After the action of the phase-gate  $S^{\alpha_j}$ and Hadamard-gate $H$ on the ancilla $A_1$, the density matrix is expressed as follows:
\begin{align}
    \rho^{(\text{ii})} & = \frac{1}{2}\ket{0}\bra{0}\otimes\left(\rho_j + O_j\rho_j\,O_j +i^{\alpha_j}\rho_j O_j+i^{-\alpha_j}\rho_j\, O_j\right) \nonumber \\
    &+ \frac{1}{2}\ket{1}\bra{1}\otimes\left(\rho_j + O_j\rho_j\,O_j -i^{\alpha_j}\rho_j O_j-i^{-\alpha_j}\rho_j\, O_j\right)\nonumber 
    \\
    &+ \text{off-diagonal terms in } A_1.
\end{align}
After the ancilla $A_1$ is measured (with result  $m_j = \{0,1\}$), the density matrix then becomes
\begin{align}
    \rho^{(\text{iii})} &= \mathcal{N}\,\ket{m_j}\bra{m_j}\nonumber \\ &\otimes\left(\rho_j + O_j\rho_j O_j +(-1)^{m_j}i^{\alpha_j}\rho_j\, O_j+\text{h.c.}\right),
\end{align}
where $\mathcal{N}^{-1} = \text{Tr}(\rho_j + O_j\rho_j\,O_j +(-1)^{m_j}i^{\alpha_j}\rho_j\, O_j+\text{h.c.})$.
If we then perform the dissipative time evolution using register $A_2$ from time $t_{j}$ to time $t_{j+1}$ and trace out the bath degrees of freedom, we obtain the following density matrix for the system register:
\begin{align}
    \rho_{j+1} \propto \mathcal{V}_{t_{j+1},t_j}\left(\rho_j + O_j\rho_j\,O_j +(-1)^{m_j}i^{\alpha_j}\rho_j O_j+\text{h.c.}\right),
\end{align}
which coincides with Eq.~(5) in the main text.
}

\section{Extra measurements for the two-time correlation function}
\begin{figure}[htpb!]
\begin{center}
    \includegraphics[width = 0.75\columnwidth]{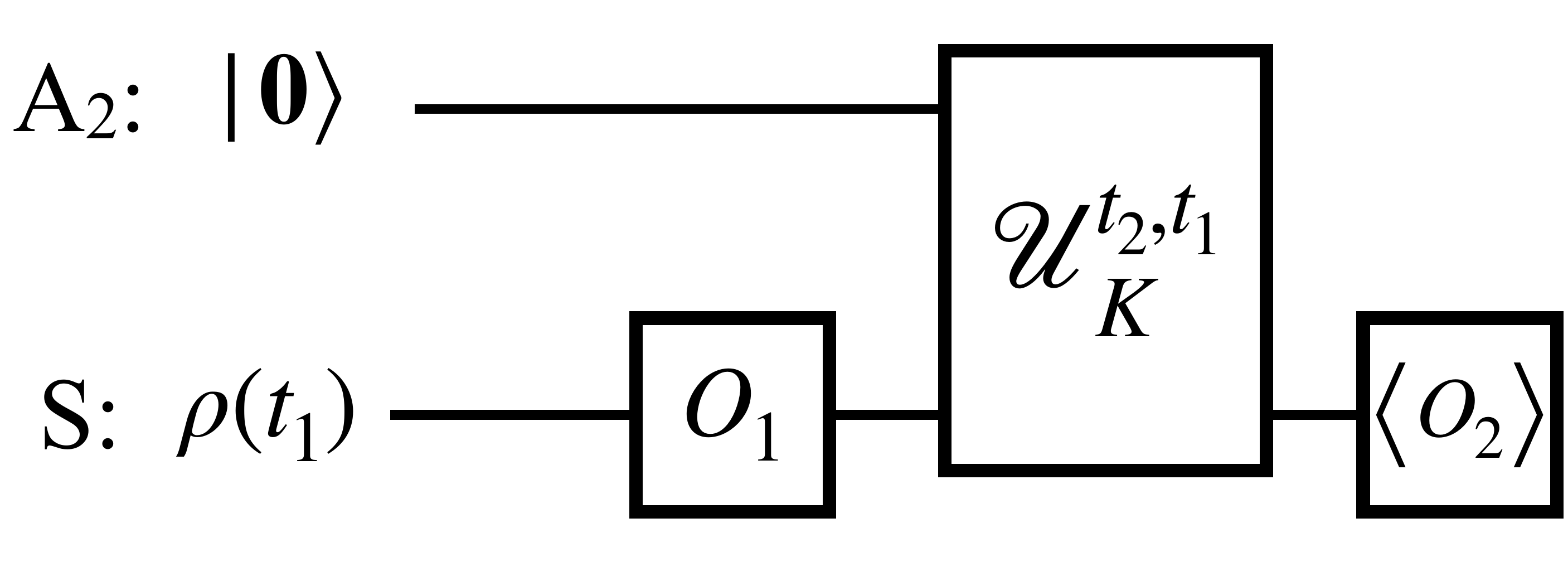}
    \caption{Quantum circuit for the measurement of the quantity $\left<O_2(t_2)\right>_{O_1}$  appearing in Eq.~(6).}\label{fig:circ_av_two_time}
\end{center}
\end{figure}

In  Figure \ref{fig:circ_av_two_time}, we show the circuit needed to obtain the information about the quantity $\left<O_2(t_2)\right>_{O_1}$ entering Eq.~(6) in the main text. We note that the extra-ancilla register $A_1$ is not needed in this case, because this quantity represents a one-time average. This quantity plus the average $\left<O_2(t_2)\right>$ multiplied times the normalisation factor $\mathcal{N} = \{2 +\big[ (-1)^{m_1}i^{\alpha_1}\left<O_1(t_1)\right> + \text{h.c.}\big]\}^{-1}$ must be subtracted from the right hand side of Eq.~(6) in order to isolate the two-time correlation function $\left<[O_1(t_1),O_2(t_2)]_\mp\right>$. Hence, in order to compute the two-time correlator, four measurements must be performed, namely the one described in Eq.~(6), and those yielding  $\left<O_2(t_2)\right>$, $\left<O_1(t_1)\right>$ and $\left<O_2(t_2)\right>_{O_1}$.

\section{Explicit calculations for the three-point correlation function}
Here we derive Eq.~(7) in the main text and give some insights about the remainder function and the additional measurements that are needed for its computation. In the derivation, we use the following abstract notation for the time evolution: $\mathcal{V}_{t_{j+1},t_j}[\rho(t_j)] = \rho(t_{j+1})$, instead of its sum representation.
\par 
The final state before the measurement in the system register in the case of $n = 3$ reads:
\begin{eqnarray}\label{eq:rho_split}
     \rho_3 &\propto& \rho_3^{(\text{I})} +  \rho_3^{(\text{II})} + \rho_3^{(\text{III})} + \left[\rho_3^{(\text{III})}\right]^\dag, 
\end{eqnarray}
where:
\begin{eqnarray}
\label{eq:rho_I}
     \rho^{(\text{I})}_3 &=& \mathcal{V}_{t_3,t_2}[ \rho_2]  \\
     \label{eq:rho_II}
     \rho^{(\text{II})}_3 &=&  \mathcal{V}_{t_3,t_2}[O_2\rho_2O_2]\\
     \label{eq:rho_III}
     \rho^{(\text{III})}_3 &=& (-1)^{m_2}\,i^{\alpha_2}\mathcal{V}_{t_3,t_2}[O_2\,\rho_2]
\end{eqnarray}

Because the time evolution preserves the trace, the normalization factor will be given by:
\begin{equation}
    \mathcal{N}_3^{-1} =2+\left[(-1)^{m_2}\,i^{\alpha_2}\text{Tr}\left[\rho_2O_2\right] + \text{h.c.}\right].
\end{equation}
There are a total of 16 terms on the RHS of Eq.~(\ref{eq:rho_split}): only four of them, those that are contained in $\rho^{(\text{III})}_3$ and its conjugate transpose,  contain the information about the three-point correlation function. The other 12 terms give rise to the remainder function $R_{\boldsymbol{\alpha}}$ when $O_3$ is finally measured. As we shall see, these 12 spurious terms contain information about two-point and one-point (averages) correlation functions. Therefore, in order to extract the three-particle correlation function we will need to perform  preliminary measurements of a few averages and two-point correlation functions that can be obtained using similar circuits as that one shown in the main text for $n = 2$.
\par
\begin{figure}[hbtp!]
    \centering
    \includegraphics[width = \columnwidth]{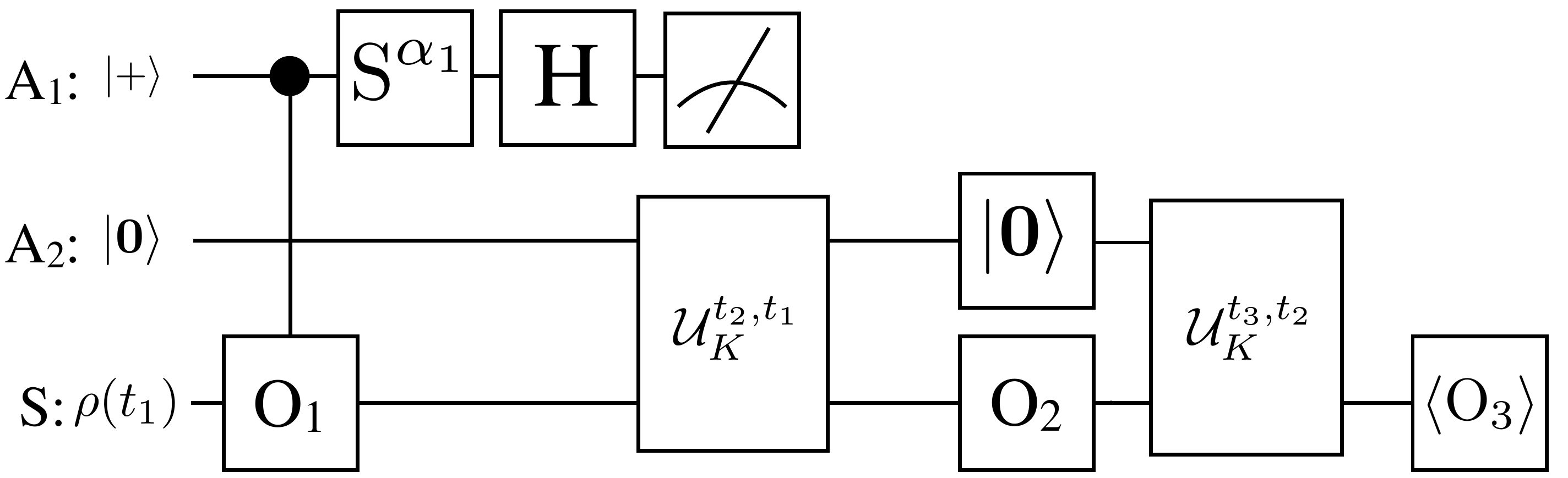}
    \caption{One  of the preliminary measurements needed in order to compute the remainder function  $R_{\boldsymbol{a}}$ in Eq.~(7).}
    \label{fig:res_func}
\end{figure}

 If we substitute $\rho_2$ from Eq.~(7) into Eq.~(\ref{eq:rho_I}) we obtain:
\begin{eqnarray}\label{eq:expl_I}
\rho_3^{(\text{I})} &=& \rho(t_3) + \mathcal V_{t_3,t_1}[O_1\rho(t_1)O_1]+ \nonumber \\
&(-1)^{m_1}&\,\mathcal{V}_{t_3,t_1}\left[i^{\alpha_1}O_1\rho_1 - i^{-\alpha_1}\rho_1O_1\right].
\end{eqnarray}
Analogously we obtain the explicit expression for Eq.~(\ref{eq:rho_II}), that reads:
\begin{eqnarray}\label{eq:expl_II}
\rho_3^{(\text{II})} &=& \mathcal{V}_{t_3,t_2}[O_2\rho(t_2)O_2] \nonumber \\
&+& \mathcal V_{t_3,t_2}\left(O_2\mathcal V_{t_2,t_1}[O_1\rho(t_1)O_1]O_2\right) \nonumber \\
&+&(-1)^{m_1}\,i^{\alpha_1}\mathcal V_{t_3,t_2}\left(O_2 \mathcal  V_{t_2,t_1}[O_1\rho_1]O_2\right) \nonumber \\
&+&(-1)^{m_1}i^{-\alpha_1} \mathcal V_{t_3,t_2}\left(O_2 \mathcal   V_{t_2,t_1}[\rho_1O_1]O_2\right).
\end{eqnarray}
It is worthwhile to note that all the terms in Eqs.~(\ref{eq:expl_I}) and (\ref{eq:expl_II}) contribute solely to the remainder function. For example, let us consider the following average that emerges from the last two terms in the RHS of Eq.~(\ref{eq:expl_II}), i.e. $i^{\alpha_1}\text{Tr}\,O_3\,\mathcal V_{t_3,t_2}\left(O_2\mathcal V_{t_2,t_1}[O_1\rho_1]O_2\right) + \text{h.c.}$:  Such a quantity can be measured using the circuit shown in Fig.~\ref{fig:res_func}, 
where the unitary operation $O_2$  acts solely on the system qubits after the time evolution from $t_1$ to $t_2$.

\par
\begin{figure}[hbtp!]
    \centering
    \includegraphics[width = \columnwidth]{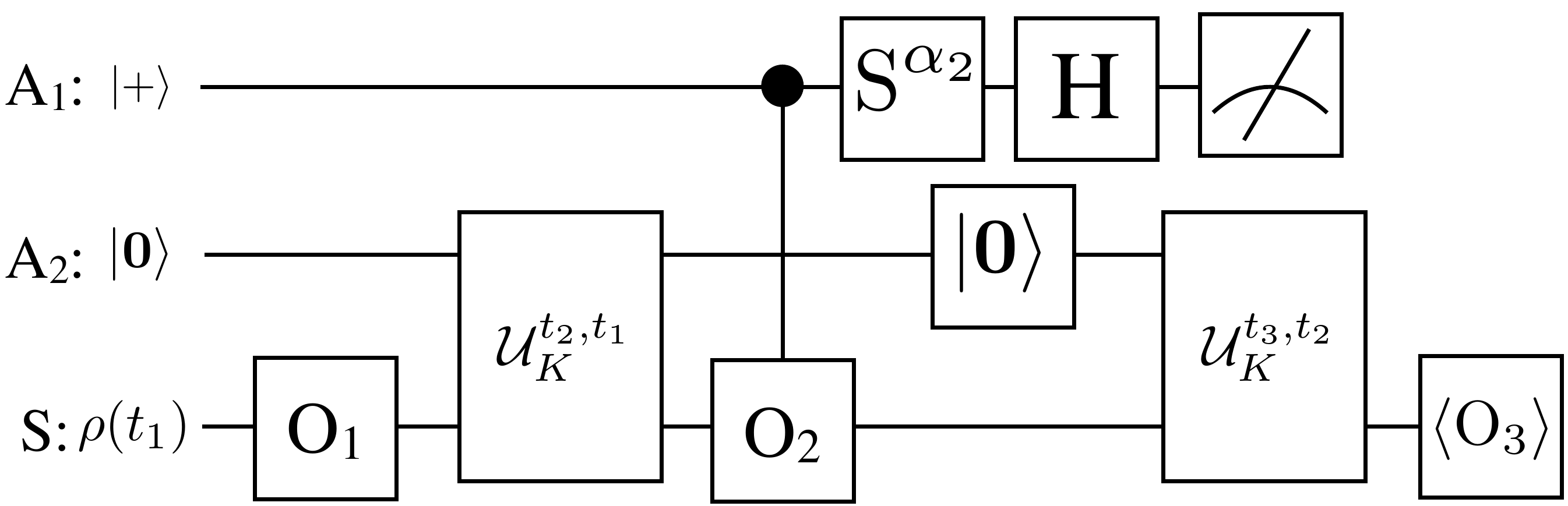}
    \caption{Quantum circuit needed in order to compute a  contribution to the remainder function   $R_{\boldsymbol{\alpha}}$  arising from $\rho^{(\text{III})}_3$ and its transpose conjugate [see Eq.~(\ref{eq:expl_III}) and the  text beneath].}
    \label{fig:res_func2}
\end{figure}

 The explicit expressions of $\rho_3^{(\text{III})}$  reads:
\begin{eqnarray}\label{eq:expl_III}
\rho_3^{(\text{III})} &=&(-1)^{m_2}\,i^{\alpha_2}\mathcal V_{t_3,t_2}[O_2\rho(t_2)]\nonumber  \\
&+&(-1)^{m_2}\,i^{\alpha_2}\mathcal V_{t_3,t_2}\left(O_2 \mathcal  V_{t_2,t_1}[O_1\rho(t_1)O_1]\right) \nonumber \\
&+&(-1)^{m_1+m_2}i^{\alpha_2+\alpha_1}\mathcal  V_{t_3,t_2}\left(O_2\mathcal V_{t_2,t_1}[O_1\rho(t_1)]\right) \nonumber \\
&+&(-1)^{m_1+m_2}i^{\alpha_2-\alpha_1}\mathcal V_{t_3,t_1}\left(O_2\mathcal  V_{t_2,t_1}[\rho(t_1)O_1]\right). \nonumber \\
\end{eqnarray}
We notice that the first two terms in the RHS of the last equation contribute to the remainder function. For example, the second term in the RHS of Eq.~(\ref{eq:expl_III}) would give rise to the following average $i^{\alpha_2}\,\text{Tr}\,O_3\mathcal V_{t_3,t_2}\left(O_2\mathcal    V_{t_2,t_1}[O_1\rho(t_1)O_1]\right) + \text{h.c.}$ that could be measure using the circuit displayed in Fig.~\ref{fig:res_func2}.
\par Therefore, by singling out the last two terms of Eq.~(\ref{eq:expl_III}), adding them up to their Hermitian conjugates, multiplying the results times $O_3$ and computing the trace, we obtain the following three-point correlation function:
\begin{eqnarray}
C^{\boldsymbol{\alpha}}_{t_1,t_2,t_3} &=& i^{\alpha_2+\alpha_1}\overbrace{\text{Tr}\,O_3\,\mathcal V_{t_3,t_2}\left(O_2\mathcal V_{t_2,t_1}[O_1\rho(t_1)]\right)}^{\left<\text{O}_3(t_3)\text{O}_2(t_2)\text{O}_1(t_1)\right>} \nonumber \\
& +&i^{\alpha_2-\alpha_1}\overbrace{\text{Tr}\,O_3\,\mathcal V_{t_3,t_2}\left(O_2\mathcal  V_{t_2,t_1}[\rho(t_1)O_1]\right)}^{\left<\text{O}_1(t_1)\text{O}_3(t_3)\text{O}_2(t_2)\right>} 
 \nonumber \\
 &+& \text{h.c.}
 \nonumber \\
\end{eqnarray}
that is the one appearing in Eq.~(7) in the main text.
\par The remainder function $R_{\boldsymbol{\alpha}}$ can be obtained in a similar way but adding up all the remaining 12 terms appearing in Eq.~(\ref{eq:rho_split})  multiplying them times $O_3$ and computing the trace. In order to extract the information about $C^{\boldsymbol{\alpha}}_{t_1,t_2,t_3}$, we will need to perform additional measurements similar to those shown in Figs. \ref{fig:res_func} and \ref{fig:res_func2}. 

\section{Specifics of the model system implemented on the quantum computer}
We model spinless electrons hopping on a one-dimensional lattice with nearest-neighbor hopping. The electrons are placed in an electric field and are connected to a fermionic bath that provides dissipation ton the system. See Refs.~\onlinecite{delre2020,rost2021}  for more details.
\subsection{Time evolution}
The time evolution of such a model  from time $t$ to $t+\Delta t$ is   governed by the infinitesimal Kraus map that can be derived from the master equation and it is given by the following set of  operators: 
 \begin{align}\label{eq:kraus_toy}
    K_0 &= 
     d^\dag_k d^{\,}_k \sqrt{1-2\,\Gamma n_F[-\epsilon_k(t)]\Delta t}\,e^{-i\epsilon_k(t) \Delta t} 
     \nonumber \\
     &+ d^{\,}_kd^\dag_k \sqrt{1-2\Gamma n_F[\epsilon_k(t)]\Delta t},\nonumber \\
    K_1 &= \sqrt{2\Gamma\,\Delta  t\,n_F[\epsilon_k(t)]}\,d^\dag_k, \\
    K_2 &= \sqrt{2\Gamma \Delta t\,n_F[-\epsilon_k(t)]}d_k^{\,},\nonumber
\end{align}
where $\epsilon_k(t) = - 2J \cos\left(k+\Omega\,t\right)$, with $J$ being the hopping amplitude, $\Omega$ the amplitude of the applied DC field, and $k$ the crystalline momentum.   $\Gamma$ measures the strength of the system-environment coupling and $n_F(x) = \left[1 + \exp(\beta\, x)\right]^{-1}$ is the Fermi-Dirac distribution with $\beta$ being the inverse of the bath temperature. The Kraus map in Eq.~(\ref{eq:kraus_toy}) can be implemented on the system qubit using an extra-ancilla qubit as shown in the main manuscript.

\subsection{Calculation of the Green's function}
The Kraus map shown in Eq.~(\ref{eq:kraus_toy}) gives the time evolution from time $t$ to time $t + \Delta t$ and it is exact in the limit of $\Delta t \to 0^+$, for this reason we refer to it as an infinitesimal map. In order to write down the Green's function analytically it is useful to derive the integrated map $V_{t,t^\prime}$ that gives the time evolution from time $t$ to $t^\prime > t$ with $t^\prime-t$ being a positive finite arbitrary number. This is equivalent to determining the Choi matrix, that \cite{andersson2007,delre2020} reads: $S_{ab}(t) = 
\sum_{r_0}^3\sum_{s = 0}^3F_{sr}(t)\text{Tr}[\sigma_r\sigma_a\sigma_s\sigma_b],
$ where the indices $a,b$  specify one of the four operators $\sigma_0 = \mathbbm{1}_{2\times 2}/\sqrt{2}$, $\sigma_1 = \sigma^x/\sqrt{2}$, $\sigma_2 =
\sigma^y/\sqrt{2}$, $\sigma_3 =
\sigma^z/\sqrt{2}$.
We can obtain a differential equation for the $F_{rs}(t)$ matrix by realizing that the master equation in the main text can be written as $\partial_t \rho = \Lambda_t(\rho)$, with $\Lambda_t$ being a linear map such that $\Lambda_t(\rho)$ is Hermitian and traceless. The equation of motion for $F$ reads: $\dot{F} = L(t)\cdot F(t)$, where the center dot indicates the matrix product, with initial condition $F(0) = \mathbbm{1}_{4\times 4}$ and with $L_{rs} = \text{Tr}[\sigma_r \Lambda_t(\sigma_s)]$ . Once the Choi matrix is obtained, we can write the time-evolution map in the Kraus representation with Kraus operators: $K_i = \sqrt{\lambda_i}\sum_{a = 0}^3 X(i)_a\sigma_a $,
where $\lambda_i$ and $X(i)$ are respectively the eigenvalues and eigenvectors of $S$.
In this way we obtain the following Kraus map: 
\begin{eqnarray}\label{eq:K_int}
K_1 &=& A_1(k,t,t^\prime)P_0+B_1(k,t,t^\prime)P_1,  \nonumber \\
K_2 &=& A_2(k,t,t^\prime)P_0+B_2(k,t,t^\prime)P_1, \nonumber \\
K_3 &=&  \alpha(k,t,t^\prime) X\,P_0,\nonumber \\
K_4 &=& \beta(k,t,t^\prime) X\,P_1,
\end{eqnarray}
where $P_0 = d^{\,}_kd^\dag_k$, $P_1 = d^\dag_kd^{\,}_{k}$, $X = \frac{1}{\sqrt{2}}(d^{\,}_k+d^{\dag}_k)$ and we have:
\begin{eqnarray}\label{eq:coeff}
A_1&=& \frac{\left(b-\sqrt{b^2+c^2+d^2}+c+i d\right) \sqrt{a-\sqrt{b^2+c^2+d^2}}}{2 (c+i
  d) \sqrt{\frac{1}{\frac{b}{\sqrt{b^2+c^2+d^2}}+1}}} \nonumber  \\
  B_1 &=& -\frac{\left(\sqrt{b^2+c^2+d^2}-b+c+i d\right) \sqrt{a-\sqrt{b^2+c^2+d^2}}}{2 (c+i
  d) \sqrt{\frac{1}{\frac{b}{\sqrt{b^2+c^2+d^2}}+1}}} \nonumber \\
  A_2 &=& \frac{\left(\sqrt{b^2+c^2+d^2}+b+c+i d\right) \sqrt{a+\sqrt{b^2+c^2+d^2}}}{2 (c+i
   d) \sqrt{\frac{1}{1-\frac{b}{\sqrt{b^2+c^2+\text d^2}}}}} \nonumber \\
   B_2 &=&\frac{\left(\sqrt{b^2+c^2+d^2}+b-c-i d\right) \sqrt{a+\sqrt{b^2+c^2+d^2}}}{2 (c+i
   d) \sqrt{\frac{1}{1-\frac{b}{\sqrt{b^2+c^2+d^2}}}}} \nonumber  \\
\end{eqnarray}
and $a = \frac{1}{2}[1 + e^{-2\Gamma(t-t^\prime)}]$, $b = \cos[f_k(t,t^\prime)]e^{-\Gamma(t-t^\prime)}$, $c =  \frac{1}{2}-n_k(t^\prime)+e^{-\Gamma(t-t^\prime)}(n_k(t^\prime)-\frac{1}{2})$ with $n_k(t) = \text{Tr}[\rho_k(t) P_1]$, $d = e^{-\Gamma(t-t^\prime)}\sin[f_k(t,t^\prime)]$, $\alpha = \sqrt{c-\frac{1}{2}(1-e^{-2\Gamma (t-t^\prime)})}$, $\beta = \sqrt{c+\frac{1}{2}(1-e^{-2\Gamma (t-t^\prime)})}$.
\par The retarded Green's function can be expressed in terms of the lesser and greater Green's functions in the following way: $G^{(R)}_k(t,t^\prime) = \theta(t-t^\prime)\left[G^>_k(t,t^\prime)-G^<_k(t,t^\prime)\right]$, where $G_k^>(t,t^\prime) = -i\left<d_k^{\,}(t)d^{\dag}_k(t^\prime)\right>=-i\,\text{Tr}\left\{X\,P_1\,V_{t,t^\prime}[X\,P_0\rho_k(t^\prime)] \right\}$ and $G_k^<(t,t^\prime) = i\left<d_k^{\dag}(t^\prime)d^{\,}_k(t)\right> = i\,\text{Tr}\left\{X\,P_1\,V_{t,t^\prime}\left[\rho_k(t^\prime)X\,P_0\right]\right\}$. 
\par A generic one-qubit density matrix can be written as $\rho(t) = \rho_\text{diag} + x(t)X+y(t)Y$, where $\rho_{\text{diag}}(t) = \text{diag}(1-n_k(t),n_k(t))$. It is easy to show that the lesser or greater Green's function does not depend on the off-diagonal elements of the density matrix. In fact:
\begin{eqnarray}
&&\text{Tr}\{XP_1 V_{t,t^\prime}[(x\, X+y\,Y)XP_0 ] \} = \nonumber \\
&&(x-i\,y)\,\text{Tr}\{XP_1 V_{t,t^\prime}[P_0 ] \}  = \nonumber \\
&&(x-i\,y)\text{Tr}\{X P_1 [(|A|_1^2 + |A_2|^2)P_0 + |\alpha|^2 P_1]\} = \nonumber \\
&&(x-i\,y)|\alpha|^2\,\text{Tr}\{XP_1\} = 0 \nonumber \,.
\end{eqnarray}
Therefore, we can write the lesser Green's function as:
\begin{eqnarray}
G^<_k(t,t^\prime) &=& i\text{Tr}\left\{V_{t,t^\prime}[\rho_\text{diag}XP_0]XP_1\right\} \nonumber   \\
&=&i\,n_k(t^\prime)\,\text{Tr}\left\{V_{t,t^\prime}[P_1\,X\,P_0]XP_1\right\},
\end{eqnarray}
and by noticing that:
\begin{eqnarray}
V_{t,t^\prime}[P_1\,X\,P_0] &=& K_1^{\,}\,P_1X P_0K_1^\dag  + K_2^{\,}P_1\,X\,P_0K_2^\dag \nonumber \\
&=&P_1\,X\,P_0\left(A_1^*B_1 + A_2^*B_2\right),
\end{eqnarray}
we  obtain:
\begin{eqnarray}\label{eq:lesser}
G^{<}_k(t,t^\prime) &=& i\,n_k(t^\prime)\left(A_1^*B_1 + A_2^*B_2\right)\mbox{Tr}\left\{XP_1\,P_1X P_0 \right\} \nonumber \\
&=&i\,n_k(t^\prime)\,e^{-\Gamma (t-t^\prime)}e^{-if_k(t,t^\prime)}
\end{eqnarray}
Analogously we obtain the following expression for the   greater Green's function:
\begin{eqnarray}\label{eq:greater}
    G^>_k(t,t^\prime) &=&-i\text{Tr}\left\{X\,P_1\,V_{t,t^\prime}\left[XP_0\,\rho_{\text{diag}}\right]\right\}
    \nonumber \\
    &=&-i(1-n_k(t^\prime))e^{-\Gamma (t-t^\prime)}e^{-if_{k}(t,t^\prime)}.
\end{eqnarray}
Let us notice that only two Kraus operators, namely $K_1$ and $K_2$ were involved in the time evolution of the operator $\rho\,d_k^{\,}$ ($d^\dag_k\,\rho$) in the expression of the lesser (greater) Green's function.
From Eqs.~(\ref{eq:lesser}) and (\ref{eq:greater}), we obtain the retarded Green's function as following:
\begin{eqnarray}
    G^{(R)}_{k}(t,t^\prime) &=& \theta(t-t^\prime)[G_k^>(t,t^\prime)-G_k^<(t,t^\prime)] \nonumber \\
    &=& -i\theta(t-t^\prime)\,e^{-\Gamma (t-t^\prime)}e^{-if_k(t,t^\prime)}.
\end{eqnarray}


\section{Hardware implementation}
The data was taken on Quantinuum's model H1-2 quantum computer, a QCCD trapped ion quantum computer that (at the time of taking this data) operated with 12 qubits. Each qubit is comprised of the 
$\left|F=0,m_F=0\right\rangle$ and $\left|F=1,m_F=1\right\rangle$ hyperfine states of a 
Yb$^{171}$ ion. Memory errors are only routinely measured and reported at the depth-1 circuit time of ~20ms 
(the depth-1 circuit time is the average time required to execute a dense 
depth-1 circuit with random gate connectivity), at which time they are in the 
low $10^{4}$'s per qubit. Given that two qubit gate errors were 
approximately $2-3\times 10^{3}$, we are confident that memory errors made a 
marginal contribution to the overall circuit error.

\section{Resource estimates on the number of measurements}
{
 The number of measurements needed to attain a given precision $\epsilon$ is proportional to $\epsilon^2$ or equivalently, the precision of our measurements scales as $1/\sqrt{n}$, which is a consequence of the Hoeffding's inequality.
The total computational cost will then be given by $n$ multiplied times the number of gates needed for state preparation (which can be roughly approximated by those contained in the time evolution). In other methods, where more than one observable must be measured \cite{mitrarai2019}  this must also be taken into account (as a multiplicative factor). 
Let us note that the cost of adding the $A_1$ ancilla is negligible in comparison to the cost of performing time evolution for long times. Moreover, assuming that $O_i$ are not entangling gates acting on $\ell$ qubits, which is a sensible choice given that many correlation functions can be expressed as a linear combination of correlation functions of strings of Pauli operators, entangling the  $A_1$ ancilla used in our scheme introduces a cost of roughly 2$\ell$ extra CNOTS. However, if we assume that time-evolution is the bottleneck of the quantum circuit and contains far more CNOTS than $\ell$, then the extra complexity introduced by entangling the ancilla with the system is negligible.}

\bibliography{ref_bibilio}